\documentclass[twocolumn,aps,prc,showpacs,superscriptaddress,preprintnumbers,floatfix,nofootinbib]{revtex4}
\usepackage{epsfig,graphics}
\usepackage{graphicx}
\usepackage{dcolumn}
\usepackage{bm}
\usepackage{amsmath}
\usepackage[usenames]{color}
\usepackage{ulem} 
\usepackage[colorlinks,linkcolor=blue,urlcolor=blue,citecolor=blue,dvipdfm]{hyperref}

\usepackage{CJK}
\begin{document}

\title{ Anisotropic flow and flow fluctuations for Au + Au at  $\sqrt{s_{NN}}$ = 200 GeV  in a multiphase transport model }

\author{ L. Ma}
\affiliation{Shanghai Institute of Applied Physics, Chinese
Academy of Sciences, Shanghai 201800, China}
 \affiliation{University of  Chinese Academy of Sciences, Beijing 100049, China}
\author{G. L. Ma}
\affiliation{Shanghai Institute of Applied Physics, Chinese
Academy of Sciences, Shanghai 201800, China}
\author{Y. G. Ma}
\thanks{ Corresponding author. Email: ygma@sinap.ac.cn}
\affiliation{Shanghai Institute of Applied Physics, Chinese
Academy of Sciences, Shanghai 201800, China}
\affiliation{Shanghai Tech University, Shanghai 200031, China}

\date{\today}

\begin{abstract}

Anisotropic flow coefficients and their fluctuations are investigated for Au+Au collisions at center of mass energy $\sqrt{s_{NN}}$ = 200 GeV by using a multi-phase transport model with string melting scenario. Experimental results of azimuthal anisotropies by means of the two- and four-particle cumulants are generally well reproduced by the model including both parton cascade and hadronic rescatterings. Event-by-event treatments of the harmonic flow coefficients $v_n$ (for n = 2, 3 and 4) are performed, in which event distributions of $v_n$ for different orders are  consistent with Gaussian shapes over all centrality bins. Systematic studies on centrality, transverse momentum ($p_{T}$) and pseudo-rapidity ($\eta$) dependencies of anisotropic flows and quantitative estimations of the flow fluctuations are presented. The $p_{T}$ and $\eta$ dependencies of absolute fluctuations for both $v_2$ and $v_3$ follow similar trends as their flow coefficients. Relative fluctuation of triangular flow $v_3$  is slightly centrality-dependent, which is quite different from that of elliptic flow $v_2$. It is observed that parton cascade has a large effect on the flow fluctuations, but hadronic scatterings make little contribution to the flow fluctuations, which indicates flow fluctuations are mainly modified during partonic evolution stage.
\end{abstract}


\maketitle

\section{Introduction}

An extreme hot and dense source composed of deconfined quarks and gluons (Quark Gluon Plasma) is believed to be created in high-energy heavy-ion collision~\cite{Whitepapers}. Because the pressure gradient of source is large enough to translate an initial coordinate space asymmetry to a final momentum space anisotropy, it can be experimentally measured as anisotropic flow. Anisotropic flow, as a typical collective behavior of particles emission, has been proved as a good probe to study the formed source because it can provide important information about equation of state and transport properties of the formed matter in high-energy heavy-ion collisions~\cite{Romatschke:2007mq, Song:2007ux, Drescher:2007cd, Xu:2007jv, Greco:2008fs, Luzum:2008cw}. One of the most striking experimental results ever obtained at Relativistic Heavy Ion Collider (RHIC) is strong elliptic flow (i.e. $v_{2}$), which is defined as the second harmonic coefficient of the Fourier expansion of the azimuthal distribution of the final-state particles~\cite{Ackermann:2000tr, Adler:2001nb, Adcox:2002ms, Back:2002gz}. The averaged magnitude of elliptic flow has been extensively studied as functions of centrality, transverse momentum ($p_T$), pseudo-rapidity ($\eta$), particle type as well as energy and collision systems~\cite{Adams:2003am,Back:2004zg,STAR-BES,STAR-Kumar,Tian-BES,Ko-NST}. Since hydrodynamical models have given many comparable descriptions on the measured flow~\cite{Ollitrault:1992bk,Voloshin:1994mz,Kolb:1999it,Heinz:2004ar, Kovtun:2004de} , it indicates that the hot and dense source is a nearly perfect fluid in the early period of high-energy heavy-ion collisions~\cite{Whitepapers,Romatschke:2007mq}.

In recent years, higher orders of harmonic coefficients in the Fourier expansion of azimuthal distribution
(i.e. $v_{n}$ for n = 3, 4, 5...) have attracted much more attention since they are expected to be more sensitive to the properties of Quark Gluon Plasma and equation of state. The higher even orders of harmonic coefficients, e.g. $v_{4}$ and $v_{6}$, were systematically studied experimentally, which provide very useful information about the collision dynamics and the properties of the hot dense matter in the initial stage~\cite{Chen:2004dv}. For higher odd orders of harmonic coefficients, they were ever supposed to vanish due to the source symmetry. However, the importance of fluctuations was realized that the initial fluctuations of geometry asymmetries will be transferred into the final momentum space as the system expands, which finally lead to non-zero values of the odd harmonic flow coefficients~\cite{Alver:2010gr, Ma:2010dv,Gavin}. It was found that odd orders of harmonic flows are sensitive to not only initial condition but also shear viscosity over entropy density $\eta$/s during QGP evolution stage. The studies of the odd orders of harmonic flows, especially third flow harmonic ($v_{3}$), are thus of great interests in recent years. For instance, it has been suggested that the third harmonic flow $v_{3}$ is responsible for the ridge and shoulder structures in dihadron azimuthal correlations~\cite{Adams:2005ph, Wang:2004kfa, Putschke:2007mi, Adams:2006tj,  Xu:2010du,Sorensen:2008dm,ZhuYH,Tang}. It was also found that $v_{3}$ shows its sensitivity to viscosity and initial state granularity, based on event-by-event relativistic viscous hydrodynamic simulations~\cite{Schenke:2011bn, Schenke:2010nt, Schenke:2010rr}.

The measurements of anisotropic flow fluctuations are believed to be a good access to the initial conditions, especially for its fluctuating or correlating properties. The measurements of elliptic flow fluctuation and even higher harmonic flow fluctuations on the event-by-event basis may elucidate both the system dynamics and new phenomena which occur in the early stage of collisions~\cite{Drescher:2007ax,Alver:2007qw, Agakishiev:2011eq}. The centrality and system-size dependencies of the event-by-event elliptic flow fluctuations in Au+Au collisions at $\sqrt{s_{NN}}$ = 200 GeV have been studied by the PHOBOS and the STAR experiments~\cite{Sorensen:2006nw,Alver:2006wh,Alver:2007rm}, which show a relative large fluctuation of 40$\%$ in mid-central region. The elliptic flow fluctuations are also theoretically studied with some models including initial state fluctuations~\cite{Andrade:2006yh,Petersen:2010cw, Han:2011iy}, which disclose the correlations between final flow and the initial geometry fluctuations, as well as some information about the viscosity and other properties of the hot matter created in high-energy heavy-ion collisions.

In this paper, we present a systematic study on harmonic flows and their fluctuations with a multi-phase transport (AMPT) model. The pseudo-rapidity and transverse momentum dependencies have been systematically studied, which provide a better understanding of the source fluctuation properties. The paper is organized as follows: In Sec. II, A multi-phase transport model is briefly introduced.  The results and discussions are presented in Sec. III.  In Sec. IV, a brief summary is presented.

\section{Brief description of AMPT model}

A multi-phase transport model~\cite{Lin:2004en} is a useful Monte Carlo model to investigate evolution dynamics for high-energy heavy-ion collisions. Currently, the AMPT model has two versions, i.e. the default version and string-melting version. Both of them consist of four main dynamical components: initial condition, parton cascade, hadronization, and hadronic rescatterings.

For the initial condition, the phase space distributions of minijet partons and soft string excitations are included, which are obtained from the Heavy-Ion Jet Interaction Generator model (HIJING)~\cite{Wang:1991hta} in which the Glauber model with multiple nucleon scatterings is basically used to describe the initial state of heavy-ion collisions. The multiple scatterings lead to the fluctuations in local energy density or hot spots from both soft and hard interactions which are proportional to local transverse density of participant nucleons. In the string-melting version, both excited strings and minijet partons are melt into partons. However all partons only consist of minijet partons in the default version. Scatterings among partons are then treated according to Zhang's Parton Cascade (ZPC) model~\cite{Zhang:1997ej} which includes only two-body elastic scatterings with a cross section obtained by the following equation,
\begin{equation}
\frac{d\sigma}{dt}=\frac{9\pi\alpha^{2}_{s}}{2}(1+\frac{\mu^{2}}{s})\frac{1}{(t-\mu^{2})^{2}},
\label{q1}
\end{equation}
where $\alpha_{s}$ = 0.47 is the strong coupling constant, $s$ and $t$ are the usual Mandelstam variables and $\mu$ is the screening mass. When all partons stop interacting with each other, a simple quark coalescence model is used to combine partons into hadrons for the string-melting version. However, the Lund fragmentation is implemented for the hadronization in the default version. Thus partonic matter is then turned into hadronic matter and the hadronic interactions are subsequently simulated by using A Relativistic Transport (ART) model which includes both elastic and inelastic scatterings for baryon-baryon, baryon-meson and meson-meson interactions~\cite{Li:1995pra}.

For some collective phenomena in high-energy heavy-ion collisions, it is found that the string-melting version is much more appropriate than the default version with the help of a large parton interaction cross section~\cite{Chen:2004dv,Lin:2004en,Chen:2006ub, Zhang:2005ni, Ma:2011uma}.  Therefore, we choose the version of AMPT model with string-melting mechanism  with the parton cross section of 10 mb or 3 mb to simulate Au+Au collisions at 200 GeV in this work. We focus on the partonic and hadronic effects on flow and flow fluctuation. For this analysis, we divide the AMPT events into different centrality bins, as described in Table~\ref{table1}, where the mean values of participant nucleons $\left\langle{N_{P}}\right\rangle$ and corresponding impact parameter ranges for each centrality bin are shown.  Before we present further results, we should mention that in most calculation results which are shown in this work except Figure 7, 10 mb parton cross section was always used.

\begin{table}[htbp]
\caption{ The mean values of participant nucleons $\left\langle{N_{P}}\right\rangle$ and impact parameter intervals corresponding to different centrality bins for Au+Au collisions at 200 GeV in this work. }
\label{table1}
\centering
\begin{tabular}{p{60pt}p{100pt}p{60pt}}
\hline
\hline
Centrality & Impact parameter (fm) & $\left\langle{N_{P}}\right\rangle$\\
\hline
 0$\%$ - 10$\%$ & 0.00 - 4.42      & 345.85$\pm$0.10 \\
10$\%$ - 20$\%$ & 4.42 - 6.25      & 263.45$\pm$0.10 \\
20$\%$ - 30$\%$ & 6.25 - 7.65      & 198.20$\pm$0.05 \\
30$\%$ - 40$\%$ & 7.65 - 8.83      & 146.85$\pm$0.10 \\
40$\%$ - 50$\%$ & 8.83 - 9.88      & 106.10$\pm$0.08 \\
50$\%$ - 60$\%$ & 9.88 - 10.82     & 73.80$\pm$0.10 \\
60$\%$ - 70$\%$ & 10.82 - 11.68    & 48.80$\pm$0.10 \\
70$\%$ - 80$\%$ & 11.68 - 12.50    & 30.60$\pm$0.03 \\
\hline
\hline
\end{tabular}
\end{table}

\section{Results and Discussion}

\subsection{Anisotropic flow coefficients $v_{n}$ and $v_{n}$ fluctuations}

The collectivity in high-energy heavy-ion collisions can be measured through final particle azimuthal anisotropy~\cite{Ollitrault:1992bk}. The anisotropy coefficients are generally obtained from Fourier expansion of final particle azimuthal distribution~\cite{Poskanzer:1998yz,Voloshin:1994mz}, i.e.

\begin{equation}
E\frac{d^{3}N}{d^{3}p}=\frac{1}{2\pi}\frac{d^{2}N}{p_{T}dp_{T}dy}(1+\sum_{i=1}^{N}2v_{n}cos[n(\phi-\psi_{RP})]),
\label{q2}
\end{equation}
where $E$ is the energy of the produced particle, $p_{T}$ is the transverse momentum, $y$ is the rapidity, $\phi$ represents the azimuthal angle of particle and $\psi_{RP}$ is the reaction plane angle.
The Fourier coefficients $v_{n}$(n=1,2,3..) are typically used to characterize the different orders of azimuthal anisotropies with the form

\begin{equation}
v_{n} = \langle cos (n[\phi- \psi_{RP}])\rangle ,
\label{q3}
\end{equation}
where the bracket $\langle \rangle$ denotes statistical averaging over particles and events. In the AMPT model, reaction plane angle $\psi_{RP}$ is assigned to be zero and the flow harmonic coefficients can be written as $v_{n} = \langle cos (n\phi)\rangle $ (for n=1, 2, 3). Harmonic flow $v_{n}$ can also be calculated with respect to the participant plane angle $\psi_{n}\left\{P\right\}$ under participant coordinate system~\cite{Voloshin:2007pc} instead of reaction plane angle $\psi_{RP}$. The participant plane is defined by the principal axis of the participant zone in the following equation

\begin{equation}
\psi_{n}\left\{P\right\}  = \frac{1}{n}\left[ \arctan\frac{\left\langle {r^n \sin
(n\varphi)} \right\rangle}{\left\langle {r^n \cos (n\varphi)}
\right\rangle} + \pi \right],
 \label{q4}
\end{equation}
where $n$ denotes the $n$th-order participant plane,  $r$ and $\varphi$ are the coordinate position and azimuthal angle of each parton in AMPT initial state and the average $\langle \cdots\rangle$ denotes density weighting.  Harmonic flow coefficients with respect to participant plane are defined as
\begin{equation}
v_{n}\left\{P\right\} = \left\langle cos[n(\phi-\psi_{n}\left\{P\right\})] \right\rangle.
\label{q5}
\end{equation}

The above method for the calculation of $v_n$ is referred to as participant plane method which has been popularly used for flow calculations in different models~\cite{DerradideSouza:2011rp}. Because the participant plane angle defines the azimuthal angle of the plane by constructing initial energy distribution in coordinate space, $v_{n}\left\{P\right\}$ with event-by-event fluctuation effects included is more reasonable to compare with experimental data. In Figure~\ref{f1}, we show the event-by-event distributions of $v_{n}\left\{P\right\}$ (n=2,3,4) for all charged particles within mid-rapidity for four different centrality bins in Au+Au collisions at 200 GeV, where solid curves are Gaussian fittings. The $v_{n}\left\{P\right\}$(n=2,3,4) distributions are consistent with Gaussian shapes for all the orders and all the centrality bins from the AMPT simulations.

\begin{figure*}[htbp]
\centering
\resizebox{17.2cm}{!}{\includegraphics[bb=13 3 542 369,clip]{fig1_v2dis.eps}
\includegraphics[bb=8 9 538 376,clip]{fig1_v3dis.eps}
\includegraphics[bb=0 9 527 392,clip]{fig1_v4dis.eps}
}
\caption{(Color online) Distributions of $v_{n}\left\{P\right\}$ [n=2 (a), 3 (b), 4 (c)]  for different centrality bins in Au+Au collisions at 200 GeV from the AMPT simulations. }
\label{f1}
\end{figure*}

Several methods and techniques have been developed to estimate the flow coefficients experimentally~\cite{Poskanzer:1998yz, Voloshin:2008dg,Borghini:2001vi, Bilandzic:2010jr}, such as the event plane method that is reconstructed within mid-rapidity ~\cite{Poskanzer:1998yz} and FTPC event plane method that uses forward- or backward- going tracks in the large rapidity window to determine the event plane. These methods have been widely used in experimental analysis of flow, though it is still argued for them to have some disadvantages~\cite{Luzum:2012da,Xiao:2012uw}. On the other hand, multi-particle correlation method (or cumulant method) has successfully quantified the harmonic flow coefficients, without requiring the reaction or participant plane~\cite{Borghini:2001vi,Borghini:2000sa, Adcox:2002ms, Adler:2002pu,WangJ}.  The contribution of non-flow correlations from lower order correlations can be effectively removed with multi-particle correlation. The cumulant method is expected to partially eliminate detector effects, because it is insensitive to detector acceptance.

A Q-cumulant or direct cumulant method, which calculates cumulants without using multiple loops over tracks and generating functions, has been developed for flow analysis~\cite{Bilandzic:2010jr}. The Q-cumulant method calculates flow coefficients $v_{n}$ (n=2,3,4..) directly from particle correlations with a flow vector defined as

\begin{equation}
Q_{n}=\sum_{i=1}^{M}e^{in\phi_{i}},
\label{q6}
\end{equation}
where $M$ is the number of particles.

The cumulants are weighted by averaging over events, which can be expressed in terms of the moments of the magnitude of the corresponding flow vector,

\begin{equation}
\begin{array}{l}
\langle \langle 2 \rangle \rangle =  \langle \langle e^{in (\phi_1-\phi_2)} \rangle \rangle = \frac{\sum_{events}(W_{2})_{i}\langle 2\rangle _{i}}{\sum_{events}(W_{2})_{i}},\\
\\
\langle \langle 4 \rangle \rangle =\langle \langle e^{in (\phi_1+\phi_2-\phi_3-\phi_4)} \rangle \rangle = \frac{\sum_{events}(W_{4})_{i}\langle 4\rangle _{i}}{\sum_{events}(W_{4})_{i}},
\label{q8}
\end{array}
\end{equation}
where the double brackets denote weighted first over the particles and then over the events for two- and four-particle correlations. The weights are the total number of combinations of two- or four-particle correlations, i.e.
\begin{equation}
W_{2}=M(M-1), W_{4}=M(M-1)(M-2)(M-3),
\label{q9}
\end{equation}
which are used to minimize the effects from multiplicity fluctuations~\cite{Bilandzic:2011PhD}.

The final two- and four-particle cumulants can be written as

\begin{equation}
\begin{array}{l}
C_{n}\left\{2\right\}=\langle \langle 2\rangle \rangle ,\\
C_{n}\left\{4\right\}=\langle \langle 4\rangle \rangle -2\langle \langle 2\rangle \rangle ^{2}.
\label{q10}
\end{array}
\end{equation}

For integral (or reference) flow coefficients, they can be estimated directly from two- and four-particle cumulants by the following equations,

\begin{equation}
v_{n}\left\{2\right\}=\sqrt{C_{n}\left\{2\right\}},
\label{q11}
\end{equation}
\begin{equation}
v_{n}\left\{4\right\}=\sqrt[4]{-C_{n}\left\{4\right\}}.
\label{q12}
\end{equation}

One can also proceed the calculations of differential flow with a new definition of vectors with particles of interest (POI) and reference particle (REP). For particles labeled as POI, we define a vector

\begin{equation}
p_{n}=\sum_{i=1}^{m_{p}}e^{in\psi_{i}}.
\label{q13}
\end{equation}

For particles labeled as both POI and REP, we define a vector

\begin{equation}
q_{n}=\sum_{i=1}^{m_{q}}e^{in\psi_{i}},
\label{q14}
\end{equation}
where $m_{p}$ and $m_{q}$ are the number of selected particles. The reduced single-event averaged two- and four-particle correlations can be formalized as $\langle 2'\rangle$ and $\langle 4'\rangle$~\cite{Bilandzic:2010jr}. Thus the two- and four-particle differential cumulants are given by,

\begin{equation}
\begin{array}{l}
d_{n}\left\{2\right\}=\langle \langle 2'\rangle \rangle ,\\
d_{n}\left\{4\right\}=\langle \langle 4'\rangle \rangle -2\langle \langle 2'\rangle \rangle \langle \langle 2\rangle \rangle.
\label{q18}
\end{array}
\end{equation}

Estimations of differential flow coefficients are expressed as:

\begin{equation}
v'_{n}\left\{2\right\}=\frac{d_{n}\left\{2\right\}}{\sqrt{C_{n}\left\{2\right\}}},
\label{q19}
\end{equation}
\begin{equation}
v'_{n}\left\{4\right\}=\frac{d_{n}\left\{4\right\}}{-C_{n}\left\{4\right\}^{3/4}}.
\label{q20}
\end{equation}

In addition, it was suggested that if the Gaussian form of flow fluctuations in the participant plane is taken with the definitions of cumulants flow~\cite{Voloshin:2007pc}, one would have

\begin{equation}
v_{n}\left\{2\right\}^{2}=\langle v_{n} \rangle ^{2}+\sigma^{2}_{v_{n}}+\delta_{n},
\label{q21}
\end{equation}
\begin{equation}
v_{n}\left\{4\right\}^{2}=\sqrt{\langle v_{n} \rangle ^{4}-2\langle v_{n} \rangle ^{2}\sigma^{2}_{v_{n}}-\sigma^{4}_{v_{n}}} \approx \langle v_{n}\rangle ^{2}-\sigma^{2}_{v_{n}},
\label{q22}
\end{equation}
where $\delta_{n}$ is the non-flow contribution from correlations not related to the reaction plane, and $\sigma_{v_{n}}$ is the absolute flow fluctuation. The approximation in Equation~(\ref{q22}) is valid when $\sigma_{v_{n}} \ll v_{n}$ and higher order moments are negligible.  In above definitions, two-particle cumulant flow contains non-flow contribution, and four-particle cumulant flow is unaffected by non-flow effect but by flow fluctuations~\cite{Bhalerao:2006tp}.

If the distribution of $v_{n}\left\{P\right\}$ is with a Gaussian shape as shown in Figure~\ref{f1}, one can take advantage of difference between Equation (\ref{q21}) and Equation(~\ref{q22}) to estimate of the total flow fluctuation in the limit of small fluctuation,

\begin{equation}
\sigma_{v_{n}}(Q)=\sqrt{\frac{v_{n}^{2}\left\{2\right\}-v_{n}^{2}\left\{4\right\}}{2}} \approx \sqrt{\sigma_{v_{n}}^{2}+\delta_{n}/2},
\label{q23}
\end{equation}

\begin{equation}
R_{v_{n}}(Q)=\sqrt{\frac{v_{n}^{2}\left\{2\right\}-v_{n}^{2}\left\{4\right\}}{v_{n}^{2}\left\{2\right\}+v_{n}^{2}\left\{4\right\}}},
\label{q24}
\end{equation}
where Equation (\ref{q23}) gives an absolute flow fluctuation estimation and Equation (\ref{q24}) presents a relative flow fluctuation estimation. It should be noted that the flow fluctuations defined with above formulas contain non-flow contributions, therefore  they can only be treated as up-limits for both absolute and relative flow fluctuations. The up-limits of elliptic flow $v_2$ fluctuations with the above definition have been studied both experimentally and theoretically so far~\cite{Sorensen:2009cz,Voloshin:2007pc}.

Non-flow correlations do have substantial effects on the measurements of flow fluctuations. Previous studies show that it is difficult to disentangle non-flow effects and flow fluctuations if without other assumptions due to initial geometry fluctuations~\cite{Ollitrault:2009ie, Voloshin:2007pc}. A recent work shows that a large pseudo-rapidity gap between particles can effectively reduce short-range non-flow contributions~\cite{Xu:2012ue}. However, the validity of the large pseudo-rapidity gap assumption can not be unambiguously applicable~\cite{Alver:2007qw}, therefore non-flow is not eliminated completely in this way and additional fluctuations from unknown sources could lead to non-ignorable contribution on the difference between $v_{n}\left\{2\right\}$ and $v_{n}\left\{4\right\}$. However, the quantity $R_{v_{n}}(Q)$ can be treated as a reasonable approximation of relative flow fluctuation, under the assumption that non-flow only takes a small portion of contribution ( $\le$ 10$\%$).

\begin{figure*}[htbp]
\centering
\resizebox{17.2cm}{!}{\includegraphics[bb=22 108 560 277,clip]{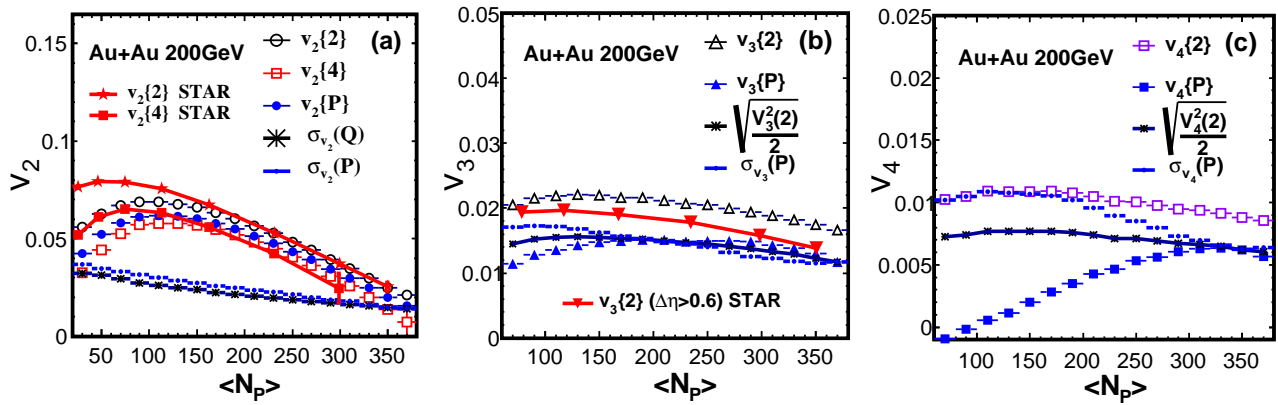}}
\caption{(Color online) Flow coefficients $v_{n}$ [n=2 (a), 3 (b), 4 (c)] for all charged particles in mid-rapidity ($|\eta|$ $\leq$ 1.0) from participant plane method and Q-cumulant method as a function of $\left\langle{N_{P}}\right\rangle$ in Au+Au collision at 200 GeV. Absolute flow fluctuation $\sigma_{v_{n}}(Q)$ (black dot line) and $\sigma_{v_{n}}(P)$ (blue dot line) are also presented, and experimental results from the STAR experiment are shown (red solid star for $v_{2}\left\{2\right\}$ and red solid square for $v_{2}\left\{4\right\}$). Experimental measurement of $v_{3}\left\{2\right\}$ with pseudorapidity gap applied is also shown for comparison (in middle panel)~\cite{Gavin}.}
\label{f2}
\end{figure*}

Since it is still uncertain if one can reliably obtain true flow fluctuations from cumulant measurements, we try a compensative way to do event-by-event treatment on $v_{n}\left\{P\right\}$ based on its standard definition in AMPT model simulations. In this way, the absolute flow fluctuation $\sigma_{v_{n}}(P)$ and the relative flow fluctuation $R_{v_{n}}(P)$ can be defined as,

\begin{equation}
\sigma_{v_{n}}(P)=\sqrt{\langle v_{n}\left\{P\right\}^{2}\rangle - \langle v_{n}\left\{P\right\} \rangle^{2}},
\label{q25}
\end{equation}

\begin{equation}
R_{v_{n}}(P)=\sqrt{\frac{\langle v_{n}\left\{P\right\}^{2}\rangle - \langle v_{n}\left\{P\right\} \rangle^{2}}{\langle v_{n}\left\{P\right\} \rangle^{2}}},
\label{q26}
\end{equation}
where brackets $\langle$ $\rangle$ denote event averaging. Figure~\ref{f2} comparatively shows the integrated flow coefficients $v_{n}$  for charged particles ($|\eta|$ $\leq$  1.0) as a function of $\left\langle{N_{P}}\right\rangle$ based on Q-cumulant method and participant plane method, as well as the absolute flow fluctuations of $\sigma_{v_{n}}(Q)$ and $\sigma_{v_{n}}(P)$ in Au+Au collisions at 200 GeV. The AMPT results for $v_{2}\left\{2\right\}$ and $v_{2}\left\{4\right\}$ are in good agreement with the STAR experimental results~\cite{Adams:2004bi} from mid-central to central Au+Au collisions. The absolute elliptic flow fluctuations $\sigma_{v_{2}}$ monotonously decrease with $N_{P}$. The standard deviation of participant flow value $\sigma_{v_{2}}(P)$ overestimates flow fluctuation $\sigma_{v_{2}}(Q)$ which contains some additional non-flow effect. It is found that $v_{2}$ is dominated by its fluctuation for the most central or the most peripheral Au+Au collisions. For higher order harmonics especially the third harmonic, the fourth power of $v_{3}\left\{4\right\}$ is almost consistent with zero within the error range in the model results which is quite consistent with the recent experiment results~\cite{Adamczyk:2013waa}. By assuming $v_{3}\left\{4\right\}$ equals to zero,$\sqrt{\frac{v_{3}^{2}\left\{2\right\}}{2}}$ is presented here instead of the complete form of Equation(~\ref{q23}) simply for comparisons. We can see for higher order harmonics $v_{3}$ and $v_{4}$,  the results from Q-cumulant method show weak centrality dependencies which is similar to the LHC measurements~\cite{ALICE:2011ab}. By comparing $v_{3}\left\{P\right\}$ and $\sigma_{v_{3}}(P)$, it is clearly seen that triangular flow mainly comes from fluctuation. From the AMPT calculations, $v_{4}\left\{P\right\}$ shows obvious centrality dependence which falls to almost zero for the  most peripheral collisions. $\sigma_{v_{4}}(P)$ only has the similar magnitude with $v_{4}\left\{P\right\}$ only for the most central events, but with larger $\sigma_{v_{4}}(P)$ than $v_{4}\left\{P\right\}$ for non-central collisions. For $v_{3}\left\{4\right\}$ and $v_{4}\left\{4\right\}$ not shown here, we found that they are consistent with zero within the errors, which is similar to the STAR preliminary measurements~\cite{Sorensen:2011fb}.

It was suggested that $v_{n}\left\{2\right\}$ should be likely equal to $v_{n}\left\{P\right\}$~\cite{Voloshin:2007pc}. For this reason, $\sigma_{v_{n}}$(P) should be smaller than $\sigma_{v_{n}}(Q)$ under positive non-flow assumption, as the difference between $\sigma_{v_{n}}(P)$ and $\sigma_{v_{n}}(Q)$ is totally caused by non-flow effect. But as shown in Figure~\ref{f2}, $\sigma_{v_{n}}(P)$ is slightly larger than $\sigma_{v_{n}}(Q)$ over all the centrality ranges which indicates that $\sigma_{v_{n}}(Q)$ is not exactly the standard deviation of $v_{n}\left\{P\right\}$ defined by $\sigma_{v_{n}}(P)$. However, it does little effect if we only make magnitude estimate of flow fluctuation and study the trends of flow coefficients and their fluctuations. Nevertheless, $\sigma_{v_{2}}(P)$ can give general estimation of $v_2$ fluctuations in magnitude, and $\sigma_{v_{n}}(P)$ can be used to study the fluctuations of higher harmonic coefficients in the similar way.

\subsection{ $v_n$ fluctuations as functions of transverse momentum and pseudo-rapidity }

It is also important to study the transverse momentum ($p_{T}$) dependence of anisotropic flow from cumulant method and make comparisons with the results from participant plane method. Figure ~\ref{f3} shows the differential $v_{2}$ and $v_{3}$ obtained with the Q-cumulant method and participant plane method, in comparison with two-particle $v_{2}\left\{2\right\}$ and four-particle $v_{2}\left\{4\right\}$ from the RHIC experiments~\cite{Afanasiev:2009wq,Snellings:2011sz}. It is observed that the AMPT results of $v_{2}\left\{4\right\}$ is consistent with the experimental data, but $v_{2}\left\{2\right\}$ is slightly larger than the experimental data. Absolute flow fluctuations of $\sigma_{v_{2}}(Q)$ and $\sigma_{v_{3}}(P)$ are also shown. It is found that absolute $v_2$ fluctuation shows similar transverse momentum and centrality dependencies as $v_{2}$. For the most central Au+Au collisions (0-10\%), $v_{2}$ is mainly dominated by its absolute fluctuation. On the other hand, $v_{3}$ has very weak centrality dependence, which differs significantly from that of $v_{2}$. It is clearly seen that absolute fluctuation of $v_{3}$ has a similar magnitude as $v_{3}$, which is consistent with the fact that  the origin of triangular flow is due to initial fluctuations.

\begin{figure}[htbp]
\centering
\resizebox{8.6cm}{!}{\includegraphics[bb=12 24 494 346,clip]{fig3_v2pt.eps}}
\resizebox{8.6cm}{!}{\includegraphics[bb=10 20 491 343,clip]{fig3_v3pt.eps}}
\caption{(Color online) Differential $v_{2}$ (Upper six panels) and $v_{3}$ (Lower six panels) as a function of transverse momentum for charged particles in mid-rapidity for six centrality bins in Au+Au collision at 200 GeV. $v_{2}\left\{2\right\}$, $v_{2}\left\{4\right\}$ are from Q-cumulant methods, and $v_{2}\left\{P\right\}$ is from participant plane method. Blacks Dots with grey shadow areas show the magnitudes of absolute flow fluctuation. The open and solid star symbols represents experimental results for two-particle $v_{2}\left\{2\right\}$ and four-particle $v_{2}\left\{4\right\}$ respectively. The absolute $v_{3}$ fluctuation from Q-cumulant method $\sqrt{v_{3}^{2}\left\{2\right\}/2}$ and participant plane method $\sigma_{v_{3}}(P)$ are represented with grey solid squares and blue dot line separately. Experimental $v_{3}\left\{TPC\right\}$ data (read solid down-triangles) which were based on the TPC event plane method are taken from Ref.~\cite{Adamczyk:2013waa} for comparison.}
\label{f3}
\end{figure}

As pseudo-rapidity ($\eta$) dependence of anisotropic flow gives additional information about the longitudinal expansion of the created medium, we investigate the dependencies of elliptic flows $v_{2}$ and triangular flow $v_{3}$ for charged hadrons on pseudo-rapidity $\eta$ with both Q-cumulant and participant plane methods with the AMPT model. Figure~\ref{f4} shows the pseudo-rapidity dependencies of the elliptic flow and triangular flow for six centrality bins in Au+Au collisions at 200 GeV from the AMPT model simulations. $v_{2}$($\eta$) for all centrality bins are of Gaussian shape, but the magnitude varies with centrality bins. However, the magnitude of triangular flow $v_{3}$($\eta$) shows weak centrality dependence as that of $v_{3}$($p_{T}$). The absolute fluctuations, $\sigma_{v_{2}}(Q)$ and $\sigma_{v_{3}}(P)$ , show similar trends as their flow coefficients. Triangular flow fluctuation of $\sqrt{v_{3}^{2}\left\{2\right\}/2}$ from Q-cumulant method are also shown for comparison.

\begin{figure}[htbp]
\centering
\resizebox{8.6cm}{!}{\includegraphics[bb=33 35 503 428,clip]{fig4_v2eta.eps}}
\resizebox{8.6cm}{!}{\includegraphics[bb=22 10 552 420,clip]{fig4_v3eta.eps}}
\caption{(Color online) The AMPT results on $v_{2}$ (upper six panels) and $v_{3}$ (lower six panels) from Q-cumulant method and participant plane method as a function of pseudo-rapidity for six centrality bins in Au+Au collisions at 200 GeV. Experimental $v_{3}\left\{TPC\right\}$ data  are also taken from Ref.~\cite{Adamczyk:2013waa}. }
\label{f4}
\end{figure}

\begin{figure}[htbp]
\centering
\resizebox{8.6cm}{!}{\includegraphics[bb=50 0 460 320,clip]{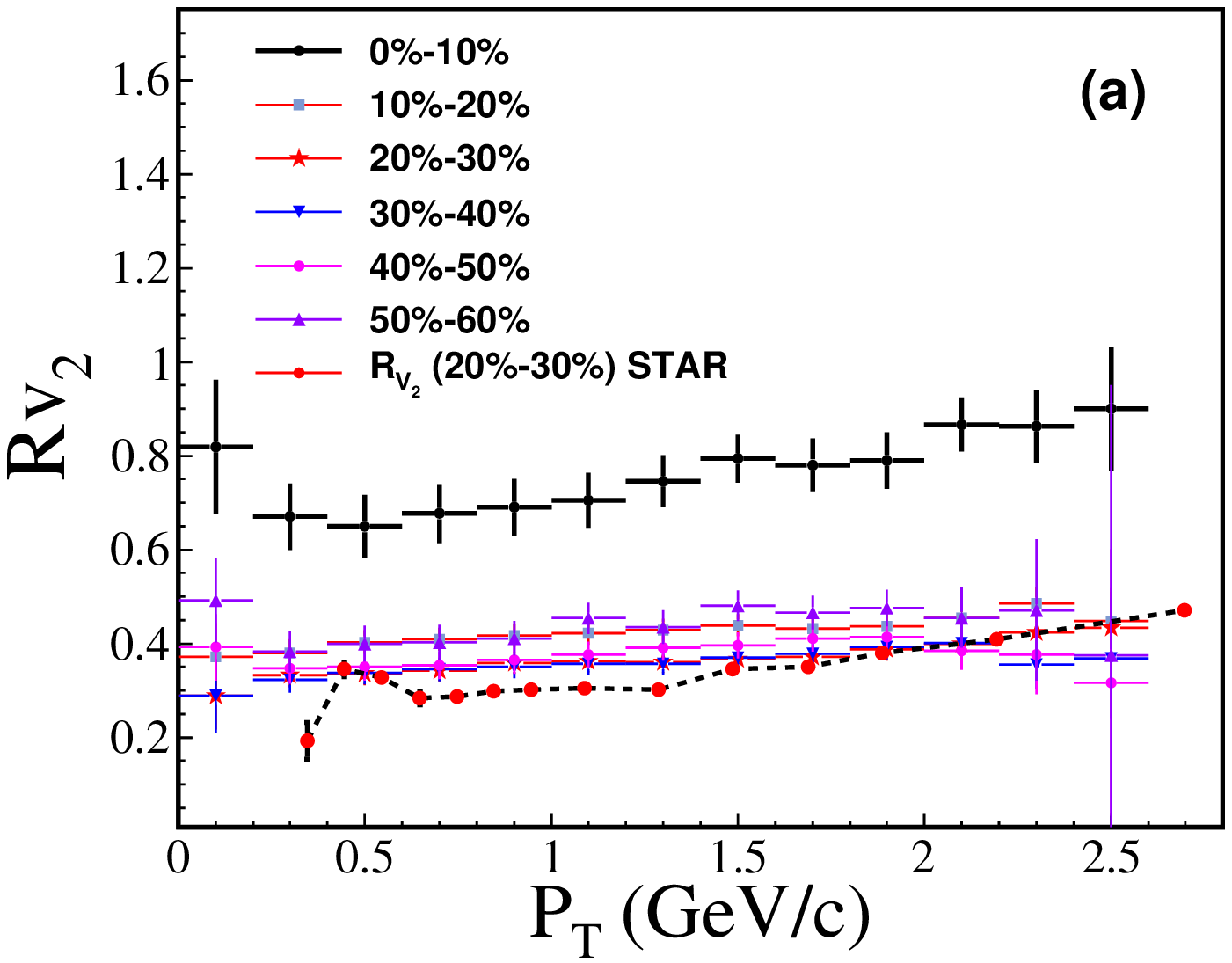}}
\resizebox{8.6cm}{!}{\includegraphics[bb=47 0 533 375,clip]{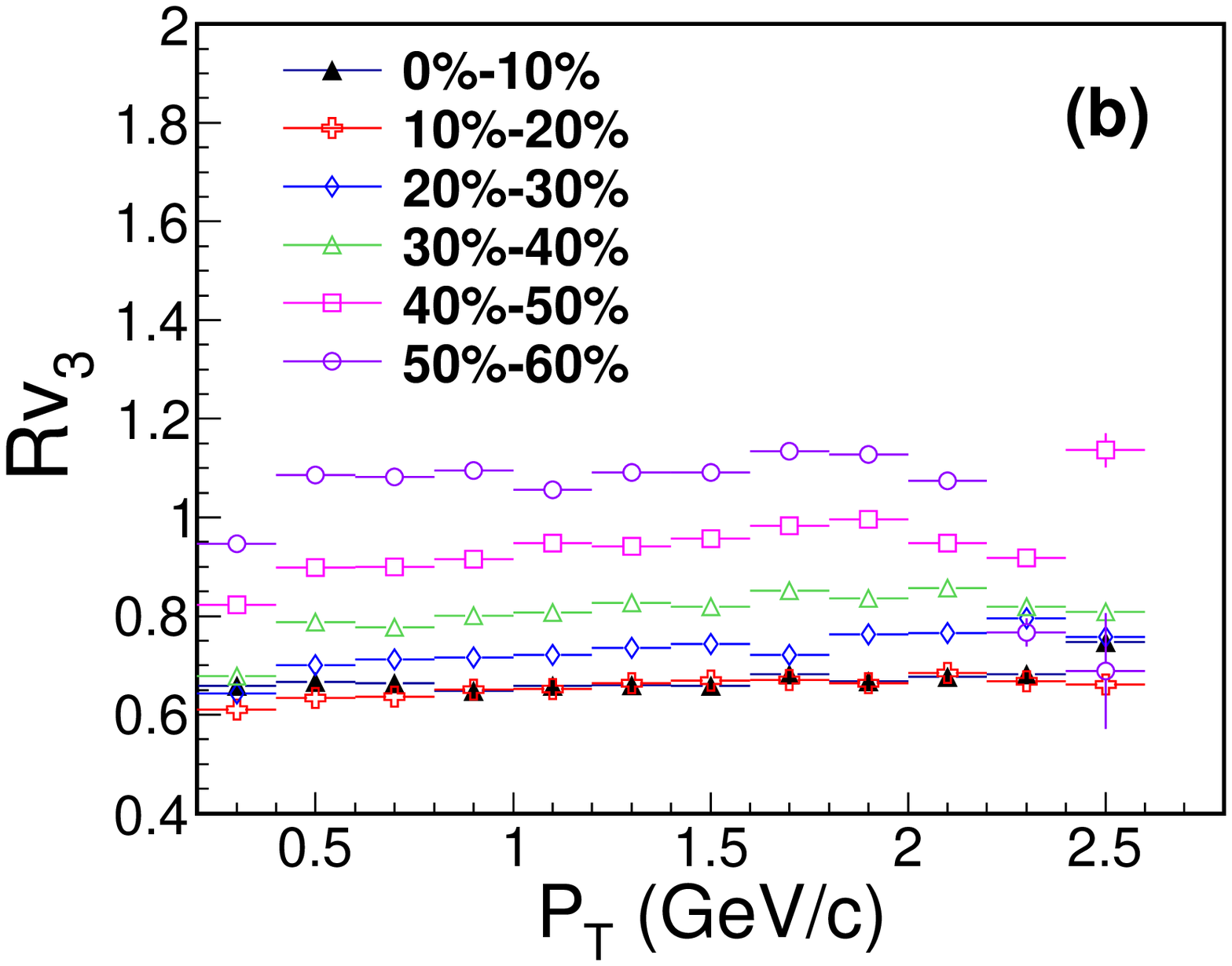}}
\caption{(Color online) The AMPT results on relative elliptic (upper panel) and triangular (lower panel) flow fluctuations as a function of transverse momentum $p_{T}$ at different centrality bins in Au+Au collisions at 200 GeV. Experimental $v_{2}$ fluctuation data  are derived from Ref.~\cite{Afanasiev:2009wq,Snellings:2011sz}.}
\label{f5}
\end{figure}

Transverse momentum dependence of relative flow fluctuation has attracted much attentions in both theoretical and experimental communities~\cite{DerradideSouza:2011rp, Abelev:2012di}. Within the framework of AMPT model, we investigate both transverse momentum and pseudo-rapidity dependencies of relative flow fluctuations for different centrality bins in Au+Au collisions at 200 GeV. In Figure~\ref{f5}, $R_{v_{2}}$($p_{T}$) shows similar trend as seen those in Pb+Pb collisions at LHC energy~\cite{Abelev:2012di}, which suggests that elliptic flow fluctuation may be controlled by a common mechanism between the two energies. $R_{v_{2}}$ for the most central collisions (0$\%$-10$\%$) shows much larger magnitude than those for other centrality bins over the whole $p_{T}$ range, which is consistent with the fact that elliptic flow is dominated by its fluctuation in the most central collisions. For $R_{v_{3}}$, it has little $p_{T}$ dependence as $R_{v_{2}}$. However, unlike $R_{v_{2}}$, $R_{v_{3}}$ shows a monotonic increasing behavior from central to peripheral collisions, which actually is driven by both centrality dependencies of $v_{3}$ and $\sigma_{v_{3}}$ together. On the other hand, it indicates the contribution of $v_{3}$ fluctuation becomes more significant in more peripheral collisions.

\begin{figure}[htbp]
\centering
\resizebox{8.6cm}{!}{\includegraphics[bb=0 0 545 383,clip]{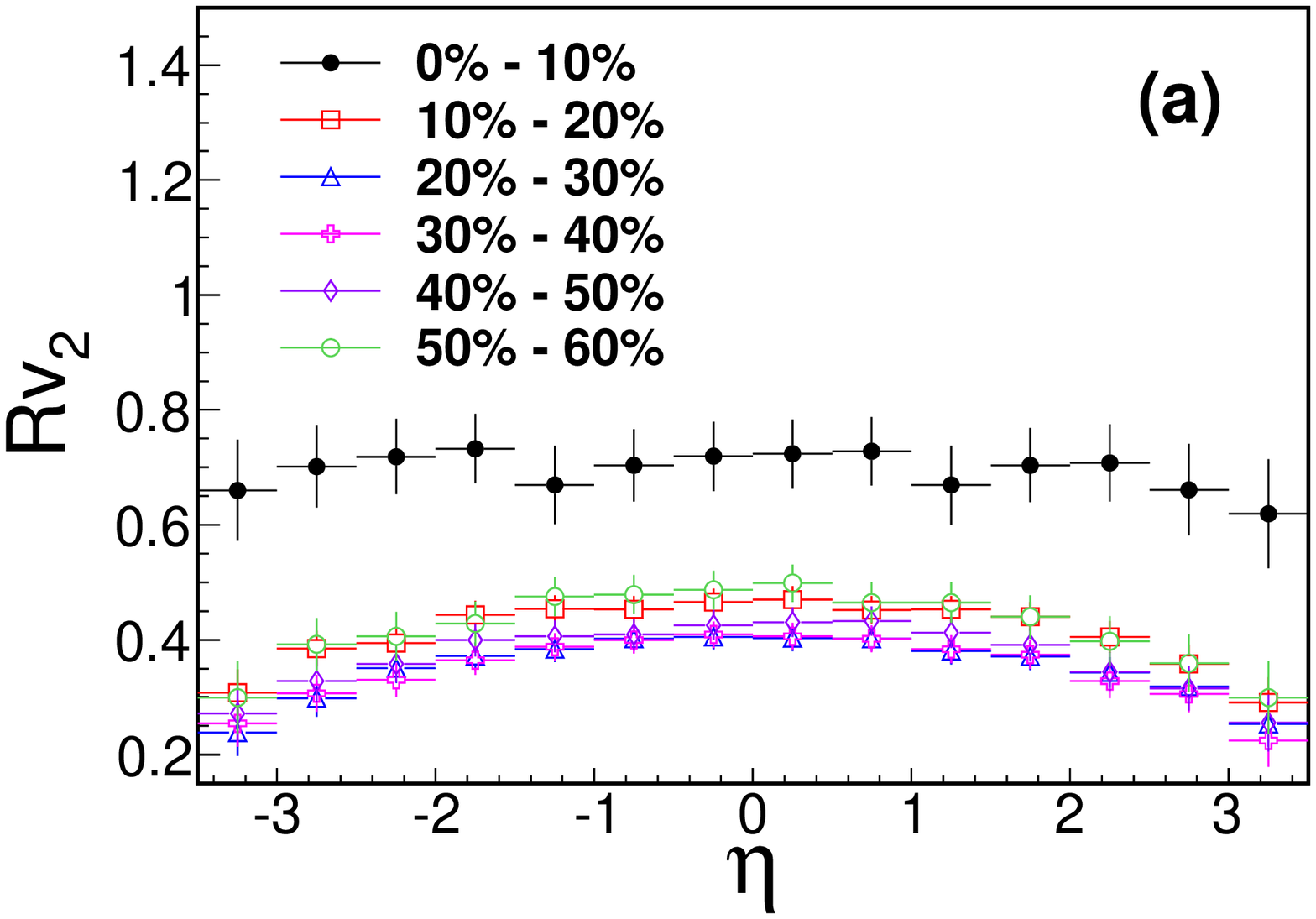}}
\resizebox{8.6cm}{!}{\includegraphics[bb=18 9 533 380,clip]{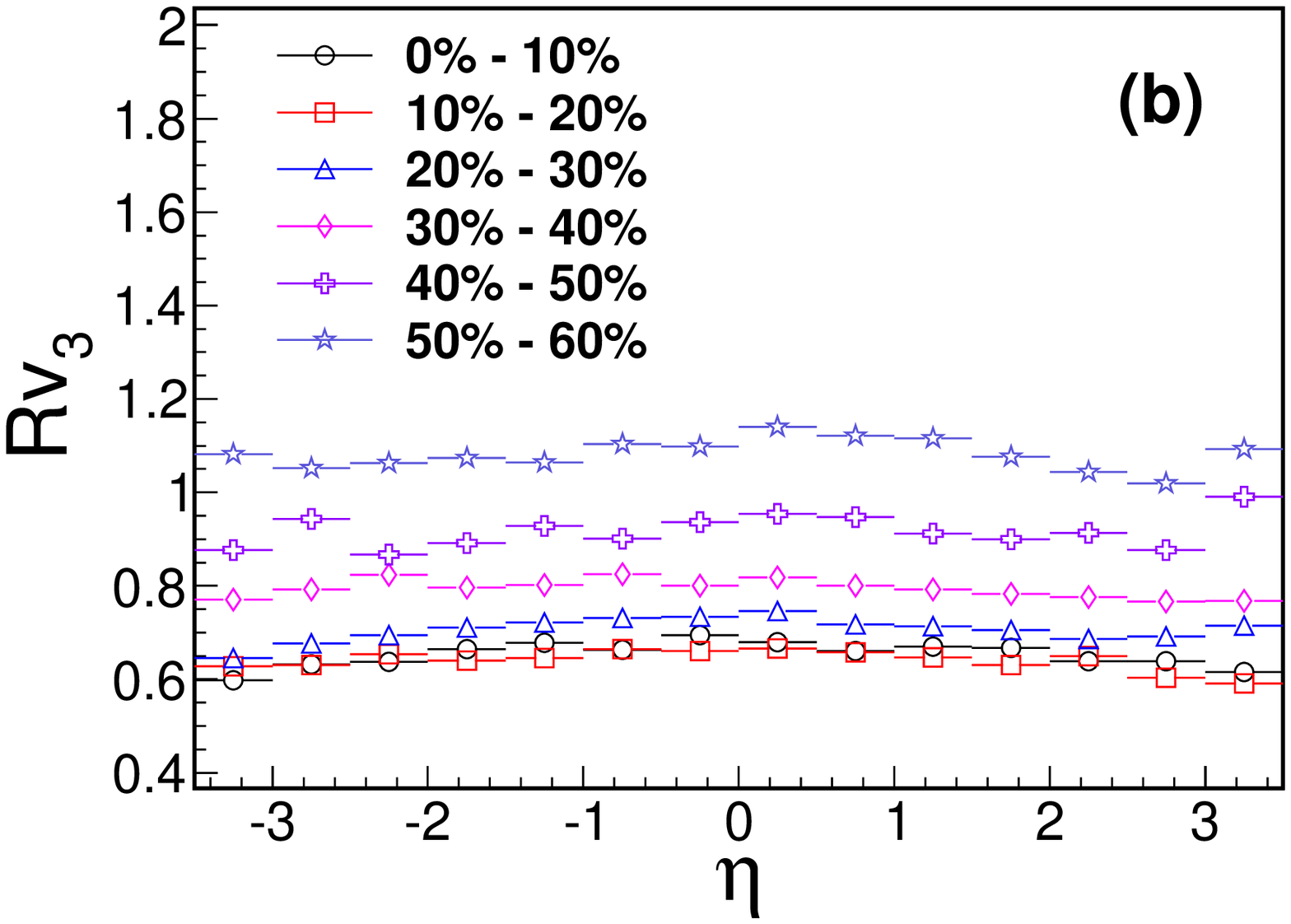}}
\caption{(Color online) AMPT results on relative elliptic (upper panel) and triangular (lower panel) flow fluctuations as a function of pseudo-rapidity $\eta$ at different centrality bins in Au+Au collisions at 200 GeV.}
\label{f6}
\end{figure}

Elliptic flow fluctuation $R_{v_{2}}$ and triangular flow fluctuation $R_{v_{3}}$ are presented  as a function of pseudo-rapidity $\eta$ in Figure.~\ref{f6}.  $R_{v_{2}}$ for the most central collisions is quite flat over the whole $\eta$ range, which is different from that for the non-central collisions in which $R_{v_{2}}$ shows slight $\eta$ dependence with a wide Gaussian shape. It is interesting that the experimental results of $v_{2}$ fluctuation versus $\eta$ shows little pseudo-rapidity dependence for Pb+Pb collision at 2.76 TeV~\cite{Hansen:2012ur}. On the other hand, $R_{v_{3}}$ shows no significant $\eta$ dependence for any centrality bin.

\subsection{ Initial partonic effect and final hadron scattering effect on flow fluctuations }

\begin{figure}[htbp]
\centering
\resizebox{9.6cm}{!}{\includegraphics{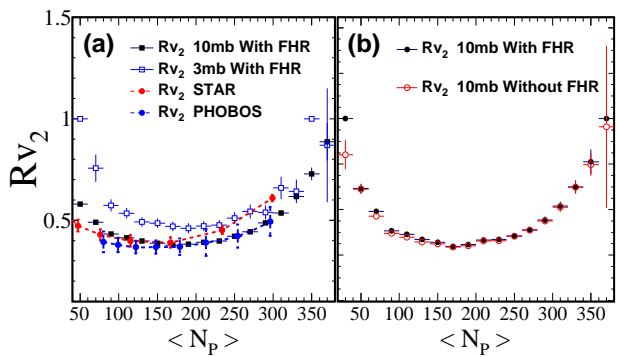}}
\caption{(Color online) The AMPT results on elliptic flow fluctuation $R_{v_{2}}$ as a function of $\left\langle{N_{P}}\right\rangle$ for Au+Au collision at 200 GeV, where panel (a) shows $R_{v_{2}}$ from AMPT simulations with parton interaction cross section of 3mb and 10mb with final hadronic rescatterings, and panel (b) shows $R_{v_{2}}$ with and without hadronic scatterings. Flow fluctuations are compared to STAR and PHOBOS data~\cite{Agakishiev:2011eq,Alver:2007qw}. The PHOBOS data errors are quoted from Ref.~\cite{BAlver:2010}.}
\label{f7}
\end{figure}

The parton interaction cross section in the AMPT model has shown a significant effect  on final flow coefficients, meanwhile final hadronic rescatterings (FHR) also have considerable influence on the magnitude of the flow coefficients~\cite{Chen:2004dv}. It is essential to investigate the effects on anisotropy fluctuations from partonic stage and hadronic stage, since it may shed light on the evolution dynamics of the source in high-energy heavy-ion collisions.  Figure~\ref{f7} shows that relative fluctuations of elliptic flow $v_{2}$ as a function of $\left\langle{N_{P}}\right\rangle$ for charged hadrons in mid-rapidity ($|\eta|$$\leq$ 1) in Au+Au collisions at 200 GeV from AMPT simulations with parton interaction cross section of 3 mb or 10 mb. Relative fluctuation $R_{v_{2}}$ with parton interaction cross section of 10 mb shows smaller values than that with parton interaction cross section of 3 mb. It indicates that large parton cross section significantly reduces relative flow fluctuation. From the quantitative comparison between the calculations and the data, 10 mb parton cross section seems have better description to the data. Figure~\ref{f7} (b) gives the comparison of $R_{v_{2}}$ between with and without final hadronic rescatterings, which suggests that final state hadron rescattering effect essentially makes little effect on relative flow fluctuation. Since $R_{v_{2}}$ is not affected by hadron rescattering process, therefore it can reflect the information about strong partonic interactions in the early partonic stage. Experimental measurements on both integral and differential flow fluctuations could serve as prospective constraints on the initial partonic stage for theoretical model simulations.

\section{Summary}

In summary, within the framework of a multi-phase transport model, we studied anisotropic flows $v_{n}$ and their flow fluctuations based on Q-cumulant and participant plane methods for charged particles at mid-rapidity in Au+Au collisions at 200 GeV. With a large parton interaction cross section and final hadronic rescatterings, experimental results of both integral and differential flows with cumulant method are generally well reproduced. Flow fluctuations defined by cumulants and standard deviation from event-by-event treatment show similar trend but slight difference in magnitude. Elliptic flow coefficient $v_2$ for the most central collisions are dominated by flow fluctuation, but triangular flow $v_3$ comes from fluctuation for all centrality bins. Flow coefficients and flow fluctuations as functions of transverse momentum and pseudo-rapidity are also investigated, which show that absolute fluctuations for differential $v_n$ (n=2, 3) follows the same trend as their flow coefficients. The parton interaction  cross section in the AMPT model calculations shows a significant effect on flow fluctuation, which indicate the modifications of flow fluctuations arise during partonic evolution stage.


Acknowledgement:  This work was supported in part by the Major State Basic Research Development Program in China under Contract No. 2014CB845400, the National Natural  Science Foundation of China under Contract Nos. 11035009, 11220101005, 10979074, 11175232, 11375251 and the Knowledge Innovation Project of Chinese Academy of Sciences under Grant No. KJCX2-EW-N01.

\end{document}